\documentclass[a4paper,pdftex]{article}

\usepackage[utf8]{inputenc} 
\usepackage[T1]{fontenc}    
\usepackage[colorlinks=true,linkcolor=black,anchorcolor=black,citecolor=black,filecolor=black,menucolor=black,runcolor=black,urlcolor=black]{hyperref}       
\usepackage{url}            
\usepackage{graphicx}       
\graphicspath{{media/}}     

\usepackage{a4wide}       

\begin{document}
\title{Variable length genetic algorithm with continuous parameters optimization of beam layout in proton therapy}
\author{F~Smekens$^{\dag}$, N~Freud$^{\dag}$, B~Sixou$^{\dag}$, G~Beslon$^{\href{https://orcid.org/0000-0001-8562-0732}{\includegraphics[scale=0.06]{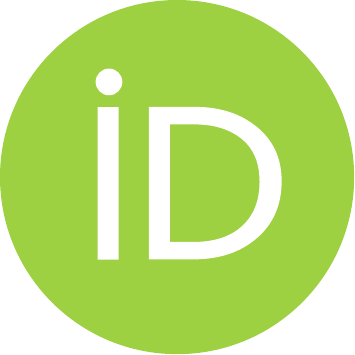}}\ddag}$, \& JM~Létang$^{\href{https://orcid.org/0000-0003-2583-782X}{\includegraphics[scale=0.06]{orcid.pdf}}\dag}$\\
  {\small $^\dag$ CREATIS, CNRS UMR5220, Inserm U1294, INSA Lyon,}\\
  {\small Universit\'e Claude Bernard Lyon 1, UJM Saint Étienne, F-69373 Lyon, France}\\
  {\small $^\ddag$ LIRIS, CNRS UMR5205, INSA-Lyon, Université Claude Bernard Lyon 1,}\\
  {\small Université Lumière Lyon 2, École Centrale de Lyon, F-69621 Villeurbanne, France}\\
}
\date{{\small Version typeset: \today}}
\maketitle

\begin{abstract}
Proton therapy is a modality in fast development. Characterized by a maximum dose deposition at the end of the proton trajectory followed by a sharp fall-off, proton beams can deliver a highly conformal dose to the tumor while sparing organs at risk and surrounding healthy tissues. New treatment planning systems based on spot scanning techniques can now propose multi-field optimization. However, in most cases, this optimization only processes the field fluences whereas the choice of ballistics (field geometry) is left to the oncologist and medical physicist.

In this work, we investigate a new optimization framework based on a genetic approach. This tool is intended to explore new irradiation schemes and to evaluate the potential of actual or future irradiation systems. We propose to optimize simultaneously the target points and beam incidence angles in a continuous manner and with a variable number of beams. No \textit{a priori} technological constraints are taken into account, \textit{i.e.}~the beam energy values, incidence directions and target points are free parameters.

The proposed algorithm is based on a modified version of classical genetic operators: mutation, crossover and selection. We use the real coding associated with random perturbations of the parameters to obtain a continuous variation of the potential solutions. We also introduce a perturbation in the exchange points of the crossover to allow variations of the number of beams. These variations are controlled by introducing a beam fluence lower limit.

In this paper, we present a complete description of the algorithm and of its behaviour in an elementary test case. The proposed method is finally assessed in a clinically-realistic test case.
\end{abstract}

\hrulefill

\smallskip
\centerline{\textsl{Correspondence should be addressed to: \texttt{jean.letang@creatis.insa-lyon.fr}}}

\hrulefill

\clearpage
\section{Introduction}

Ion beam therapy is a fast developing modality using ion beams (protons or carbon ions) to treat tumors. Characterized by a maximum dose deposition at the end of the particle trajectory (Bragg peak) followed by a sharp fall-off, ion beams can deliver a highly conformal dose to the planning target volume (PTV) while sparing the organs at risk (OAR) and the surrounding normal tissues (NT). Recently, new treatment planning systems (TPS) dedicated to spot scanning techniques have been proposed to simultaneously optimize several fields and thus improve the dose delivery \cite{Gemmel2008,Bourhaleb2008}. A typical treatment generally consists of several fractions and requires a computed tomography (CT) scan, the contouring of the tumor and of critical structures, and the inverse planning, \textit{i.e.}~the optimization of the beam parameters that make it possible to obtain the prescribed homogeneous dose inside the PTV.

In this paper, we investigate a new optimization framework based on a genetic algorithm (GA). This tool is intended to explore new irradiation schemes and to assess the potential of actual or future irradiation systems. We propose to optimize simultaneously the target points and beam incidence angles in a continuous manner and with a variable number of beams. No \textit{a priori} technological constraints are applied to the beam energy values, incidence directions and target points. The TPS is generally divided into two parts: the dose simulation engine and the optimization algorithm. Our work focuses on the optimization algorithm which has to converge towards an optimum with any dose simulation scheme.

In the spot scanning active delivery mode (intensity modulated particle therapy -- IMPT), a proton beam with a transverse Gaussian profile (typically 5--10~mm FWHM) is scanned in two dimensions using magnetic devices~\cite{Lomax1999}. The proton penetration depth in matter is adjusted by tuning the beam energy. Note that carbon ions can be used in the same way and present an higher (variable) biological effectiveness which has to be taken into account in the TPS using suitable models. The treatment plan generally combines less than $6$ fields for which the entrance channels are manually selected. For each field, an analytical method is used to cover the PTV with spots, then a gradient descent is used to adjust the beam fluences to obtain a homogeneous dose in the tumor and to limit the dose in OAR~\cite{Kramer2009}.

In the IMRT field, planning techniques have been investigated with mixed analytical and probabilistic optimization, such as genetic algorithms~\cite{Li2004,Chaomin2008} and simulated annealing~\cite{Webb2005,Hartmann2008}. In most cases, the probabilistic method only addresses the optimization of a single aspect, \textit{i.e.}~the angles, shape or weight of the beams~\cite{Li2004,Lei2008,Chaomin2008,Cotrutz2003,Li2003}, whereas the others are processed with an analytical method or based on clinical experience.

Among the existing evolutionary algorithms, Messy GA~\cite{Goldberg1993} and the state-of-the-art evolutionary strategy {CMA-ES} (Evolution Strategy with Covariance Matrix Adaptation)~\cite{Hansen2003} present very good convergence and speed properties. As Messy GA have recourse to very specific operators and {CMA-ES} to a heavy mathematical formalism, we preferred to stick to classical genetic operators (mutation, crossover and selection), which can be more easily adapted to the inverse planning problem, based on physical considerations. These operators were modified to optimize the parameters, i.e.~target points and beam incidences, in a continuous space and to allow variations of the individual genome length, \textit{i.e.}~of the number of beams. The continuous aspect of our method was introduced by using the real coding of the individuals instead of the binary coding~\cite{Janikow1991,Herrera1998,Vukovic1999} and a specific non-uniform mutation scheme~\cite{Michalewicz1992}. Genome size variations were implemented using a cut-and-splice crossover operator~\cite{Cavill2006}.

In section~\ref{sect:optimizationScheme}, the dose simulation scheme adopted in this work is succinctly described (section~\ref{ssect:doseModel}), followed by a description of the proposed GA (section~\ref{ssect:variableLengthGA}). The convergence results presented in section~\ref{sect:results} are discussed in terms of dynamics and efficiency in a simple test case (section~\ref{ssect:globalGAdynamic}) and in a clinically-realistic case (section~\ref{ssect:humanModelOptimization}).

\section{Optimization scheme}

\label{sect:optimizationScheme}

We propose to optimize a pencil beam scanning (PBS) treatment plan in protontherapy based on a GA. The beam target positions, incidence angles as well as the number of beams are optimized simultaneously. The dose distribution model is described in section~\ref{ssect:doseModel} and the different parts of the GA are detailed in section~\ref{ssect:variableLengthGA}.

\subsection{Dose calculation}

\label{ssect:doseModel}

In this study, a simplified pencil beam description is adopted to favor speed as the GA approach requires many generations (typically $10^3$) and, for each generation, many dose simulations. Every pencil beam is described as a single ray. It is worthy of note that this model does not account for the lateral straggling of ions and the Gaussian transversal profile of the pencil beams. The main steps of the dose calculation are the following:

\begin{itemize}


\item{} A beam is oriented in space by its incident direction (angles $\theta$ and $\phi$) and target position (vector $\textbf{p}$) considered as the Bragg peak position. This determines the energy of the beam. The corresponding depth-dose profile is obtained using a set of profiles in water, pre-calculated using the Monte Carlo (MC) code Geant4~\cite{Agostinelli2003,Allison2006}. The material heterogeneities are taken into account by the water equivalent path length method~\cite{Batin2008}.


\item{} Each beam is propagated in a voxel matrix using a fast raycasting method~\cite{Siddon1985,Jacobs1998,Zhao2003}. This method returns the list of voxels traversed by the beam with the corresponding path lengths. It makes it possible to calculate the dose deposited by a beam with unit fluence in every voxel along its trajectory.


\item{} A fast gradient-based method is then used to optimize the beam fluence values, so that a homogeneous dose is obtained in the PTV.

\end{itemize}

Note that the dose simulation model is independent from the GA optimization and can be replaced in a straightforward way by physically more realistic models (with increased computational cost).

\subsection{Variable length genetic algorithm}

\label{ssect:variableLengthGA}

In the GA approach, a set of individuals represented by a list of genes evolves simultaneously through the landscape, \textit{i.e.}~the solution space of the problem. The proposed method is based on the classical mutation, crossover and selection GA operators:

\begin{itemize}
\item{} Mutation modifies the genes of individuals with a probability $p_m$ (generally small).
\item{} Crossover mixes two individuals by exchanging parts of their genetic material with a probability $p_c$ (generally high).
\item{} Selection retains an individual in the next generation with a selection probability based on its fitness score.
\end{itemize}

These operators were adapted to the context of treatment planning, including (i) a continuous parameter representation (through real parameter coding), (ii) a non-uniform mutation scheme, (iii) a clinically-relevant objective function associated to an exponential ranking operator, and (iv) a variable-length individual genome, \textit{i.e.}~number of beams (through a cut-and-splice crossover).

\subsubsection{Parameter coding}

\label{sssect:parameterCoding}

In GAs, the coding of individuals (genotype) is generally different from their real representation (phenotype). The most common ways to code the genes in GAs are the binary and real coding. The binary coding represents an individual as a list of parameters (\textit{i.e.}~vectors or angles) transformed into binary chains of fixed size whereas, in the real coding, the indvidual is directly coded by the real parameter values. It has been proved~\cite{Janikow1991,Herrera1998} that real coding was able to avoid the Hamming cliffs phenomenon~\cite{Rothlauf2002} known to be present in the binary coding. Real coding is also preferable for optimization problems where the objective function is a real value~\cite{Vukovic1999}.

In what follows, an individual is defined as a set of beams, defined by their target position and incidence angles.

\subsubsection{Non-uniform mutation}
\label{sssect:mutation}

A mutation operator adapted to real parameter coding was defined. The most common way is to select a mutant parameter with a probability $p_m$ (typically about $0.01$) and to add a Gaussian noise of fixed FWHM to the parameter value. This scheme, well-suited for fixed size individuals, presents some disadvantages when considering a varying size genome. As the crossover operator makes the beam number increase, the average distance between parameter values (\textit{e.g.}~target positions in the PTV region) becomes smaller and smaller leading to an unadapted FWHM of Gaussian noise. Moreover, if the same individual is selected more than once in a generation, the small mutation probability $p_m$ is not sufficient to create a significant difference between these individuals.

To tackle these issues, a non-uniform mutation scheme was used~\cite{Michalewicz1992}. At each generation all the coding parameters are modified by adding an exponentially decreasing noise as follows:

\begin{equation}
v_p^\prime = v_p \pm \delta(\lambda_p)
\end{equation}

where $v_p$ and $v_p^\prime$ are respectively the old and new values of parameter $p$, $\delta$ is a fluctuation randomly picked in the exponentially decreasing distribution of mean $\lambda_p$. In the case of the target position, this operation is repeated for every space coordinate. An empirical rule was established to rule the evolution of $\lambda_p$: if no better solution is found during $20$ generations, all $\lambda_p$ values decrease by $10 \%$. In addition, a simple rule is defined for the non-coding parameters (see section~\ref{sssect:crossover}), specifying that the beam is randomly reinitialized, for it to contribute to the forthcoming genetic evolution.

An automatic method was developed to set the initial $\lambda_p$ values. For target positions, it is chosen equal to the mean distance between two neighbor target positions considering a uniform distribution of the initial positions in the tumor. We applied the same method for the perturbations of incidence angles $\phi$ and $\theta$ considering a uniform distribution around the tumor.

\subsubsection{Objective function and selection}
\label{sssect:objectiveFunctionAndSelection}

The purpose of the objective function is to assess the adaptation of an individual to its environment. In treatment planning, the dose distribution map of an individual is compared to a prescribed distribution segmented into regions of interest. We consider PTV, OAR and NT region types. The score is calculated using squared residuals~\cite{Schreibmann2004,Li2004} as follows:

\begin{equation}
\label{equ_fitnessFunction}
f_{obj} = \sum_{i=0}^{I} f_i
\end{equation}
with
\begin{equation}
\label{equ_quadraticFitnessFunction}
f_i = \frac{1}{N_i^\ast} \sum_{n=1}^{N_i} \delta_{i,n} (d_n - p_i)^2
\end{equation}
where $f_{obj}$ is the total score of the objective function, $I$ is the total number of regions, $f_i$ is the region score, $N_i^\ast$ a normalization factor, $n$ is the $n^{th}$ voxel in the $i^{th}$ region containing $N_i$ voxels, $d_n$ is the deposited dose, and $p_i$ is the dose prescription. $\delta_{i,n}$ is a boolean flag which is set to 1 except for OAR and NT region if $d_n \le p_i$. The aim of $f_{PTV}$ is to tend to the prescribed dose values whereas $f_{OAR}$ and $f_{NT}$ ensure the dose stays lower than a limit. The normalization factor $N_i^\ast$ is set to the number of voxels in the case of PTV and OAR. As the NT region often has many more voxels than other regions, $N_{NT}^{\ast}$ represents the number of NT voxels crossed by the beams, averaged over the individuals. To account for the varying number of beams, $N_{NT}^\ast$ is updated every $50^{th}$ generation.

Associated to the objective function, a selection operator was defined. In the classical selection scheme, individuals are randomly picked in the current generation using a probability proportional to their score. However, with this method, a highly adapted individual can easily screen the others, leading to premature convergence. Therefore, an exponential ranking selection scheme~\cite{Blickle1996} was used, in which the selection probability depends on the rank of the individual and not directly on its objective function value. This highly favors the best individuals while allowing fit (but not optimal) individuals to participate to the optimization process.

\subsubsection{Cut-and-splice crossover}
\label{sssect:crossover}

The cut-and-splice crossover~\cite{Cavill2006} is a well known operator which makes it possible to introduce perturbations in the individual size (number of beams). In contrast with the one-point crossover where a single cut point is randomly chosen for the two parents, the cut-and-splice operator generates a random cut point for each of them. For example, parents of sizes $(s_1,s_2)$ and random cut point positions $(\alpha_1,\alpha2)$ will generate children of sizes $(\alpha_1 + s_2 -\alpha_2 , \alpha_2 + s_1 - \alpha_1)$. This operator was tested but did not produce satisfactory results in the absence of any constraints on the cut point positions. To deal with this issue, a modified cut-and-splice crossover scheme was proposed:

\begin{equation}
\begin{array}{ccc}
\left\{
\begin{array}{l}
s_1 = \alpha + \beta \\
s_2 = k (\alpha + \beta)
\end{array}
\right.

&

\Rightarrow

&

\left\{
\begin{array}{l}
s_1^{\prime} = \alpha + k \beta + \delta_1 - \delta_2 \\
s_2^{\prime} = k \alpha + \beta + \delta_2 - \delta_1
\end{array}
\right.
\end{array}
\end{equation}

where $s$ is the parent size, $s^\prime$ the child size, $\alpha$ the cut position of parent 1, $\beta$ the remaining length of parent 1 and $k$ the ratio of parent sizes $s_2 / s_1$. $\delta$ is a random perturbation generated with a Gaussian distribution centered at zero. Note that cut points are always chosen between beams. The FWHM is set so that $\delta$ is quite small (FWHM between $2$ and $10$ beams) leading to a slowly increasing number of beams (because individuals with larger numbers of beams provide a better geometrical coverage of the PTV). Accordingly, the mean beam fluence slowly decreases. A lower limit of the beam fluence is applied to control the beam number. If the beam fluence is below this limit (after the fluence optimization process, see section~\ref{ssect:doseModel}), the beam is considered as non-coding (its contribution to the dose is not taken into account in the objective function). A minimum number of $2 \times 10^3$ protons per beam was chosen.

\section{Results}
\label{sect:results}

In this section, we study the global dynamics of the proposed variable length GA, i.e.~how the inclusion of the beam number in the genetic process affects the evolution (see section~\ref{ssect:globalGAdynamic}). Then the optimization of a clinically-realistic test case is fully detailed (see section~\ref{ssect:humanModelOptimization}).

\subsection{Global GA dynamic}
\label{ssect:globalGAdynamic}

The introduction of the beam number in the optimized parameters modifies the GA dynamics. The genetic operators used to control the individual size have to be tuned. These two aspects are investigated through (i) a comparison between genetic algorithms with and without operators specific to the variable length and (ii) a study of the influence of the initial beam number and crossover perturbation.

\begin{figure}[htb]
\centerline{\includegraphics[width=.4\textwidth]{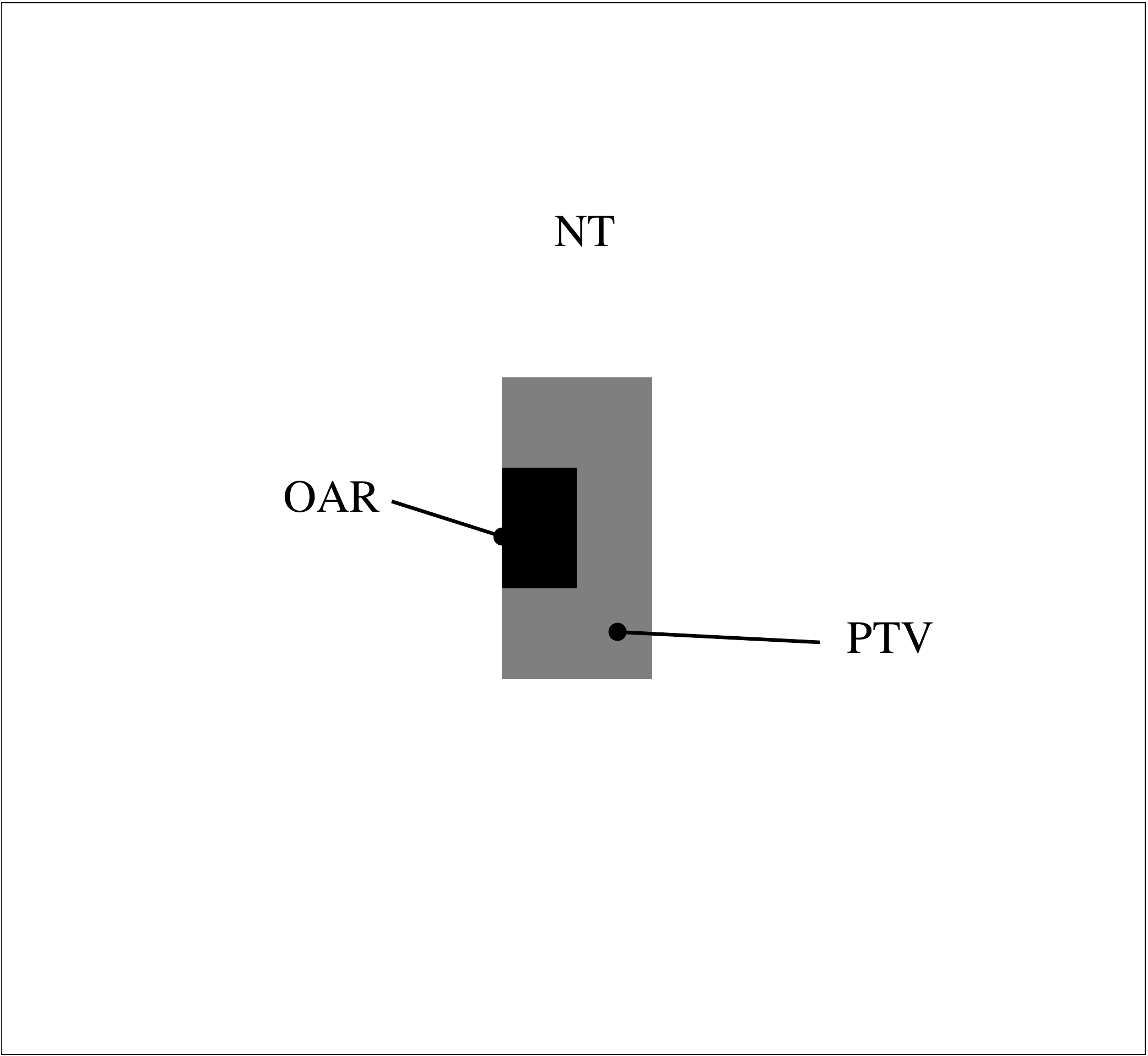}}
\caption[]{C-shaped tumor model. The volume is composed of $64 \times 64 \times 64$ cubic voxels of water with a $1$ mm voxel size.}
\label{Fig_cShapeModel}
\end{figure}

All the optimization test cases presented in this section were carried out on an artificial C-shaped tumor (figure~\ref{Fig_cShapeModel}). The volume is composed of $64 \times 64 \times 64$ of cubic voxels of water with a $1$ mm voxel size. The optimized parameters are the target position $\textbf{p}$ restricted to the tumor volume and the angle $\phi$ comprised between $0$ and $360^\circ$ (the irradiation is planar). The dose prescribed to the PTV is $4$~Gy and the limits for the OAR and NT are respectively $0$ and $1$ cGy.

\subsubsection{Effects of variable length specific operators}

The dynamics of the proposed GA is investigated by means of a comparison with two simpler algorithms in which only some of the operators presented in section~\ref{ssect:variableLengthGA} are activated. All algorithms use real coding (section~\ref{sssect:parameterCoding}) as well as the objective function and selection operator (section~\ref{sssect:objectiveFunctionAndSelection}). The two simpler algorithms are the following:

\begin{itemize}
\item The first one only uses basic operators which do not affect the number of  beams, i.e.~the classical 1-point crossover and the non-uniform mutation operator (section~\ref{sssect:mutation}) without the concept of non-coding genes.
\item The second one uses the previously proposed specific operators: cut-and-splice crossover (section~\ref{sssect:crossover}) and non-uniform mutation restricted to the random replacement of non-coding genes.
\end{itemize}

\begin{figure}[p]
\centerline{\includegraphics[width=.8\textwidth]{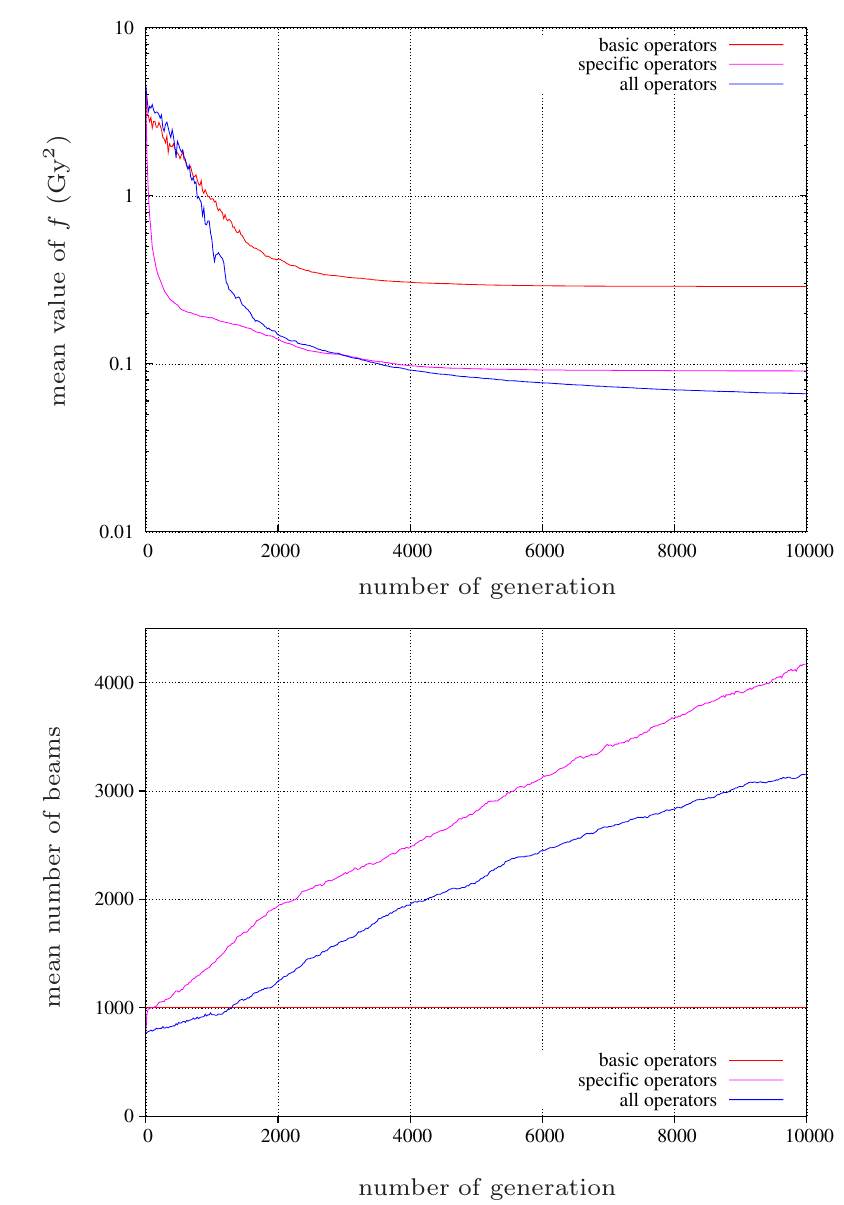}}
\caption[]{Convergence curves (top) and evolution of mean number of coding beams (bottom) in the case of the proposed algorithm (labeled 'all operators') and of the two simpler algorithms ('basic' and 'specific' operators).}
\label{Fig_specialSchemeDynamic}
\end{figure}

The population size was set to $50$ individuals, the initial beam number to $1000$ and the crossover rate to $0.5$. The convergence of the objective function and the corresponding evolution of the number of beams are represented in figure~\ref{Fig_specialSchemeDynamic}.

We can see that the three algorithms converge to different values of the objective function. With the basic operators, in spite of the fixed number of beams, the objective function (red lines) decreases sharply in the first 2000 generations and  reaches a minimum after about $6000$ generations. This behavior can be explained by the optimization of the beam layout, enhanced by the restricted number of beams. In contrast, with the specific operators, genetic diversity is generated by the constant bringing of new beams and the random replacement of non-coding beams, which leads to a very fast evolution during the $500$ first generations. However the decrease of the objective function progressively stops when the number of beams increases. It is worthy of note that the beam number continues to grow linearly without any additional benefit. In this case, the optimization of the objective function comes from the increase in the number of beams rather than the quality of the beam layout. In the case of the proposed GA ('all operators'), the evolution process displays two different regimes: (i) during the first $1000$ generations the beam layout is optimized with only slight variations in the number of beams and a drop of the mean value of the objective function; (ii) then the number of beams increases (variable length dynamics) and are successfully used to further improve the objective function by performing a local search. The final solution is better than with the specific operators alone, with a lower number of beams. The proposed algorithm is thus a satisfactory trade-off between the quality of the beam layout and the number of beams.

\subsubsection{Setting of GA parameters}

In what follows, the influence of crossover perturbation, initial number of beams and fluence limit are discussed.

\paragraph{Influence of crossover perturbation}
\label{par::crossoverPerturbation}

\begin{figure}[p]
\centerline{\includegraphics[width=.8\textwidth]{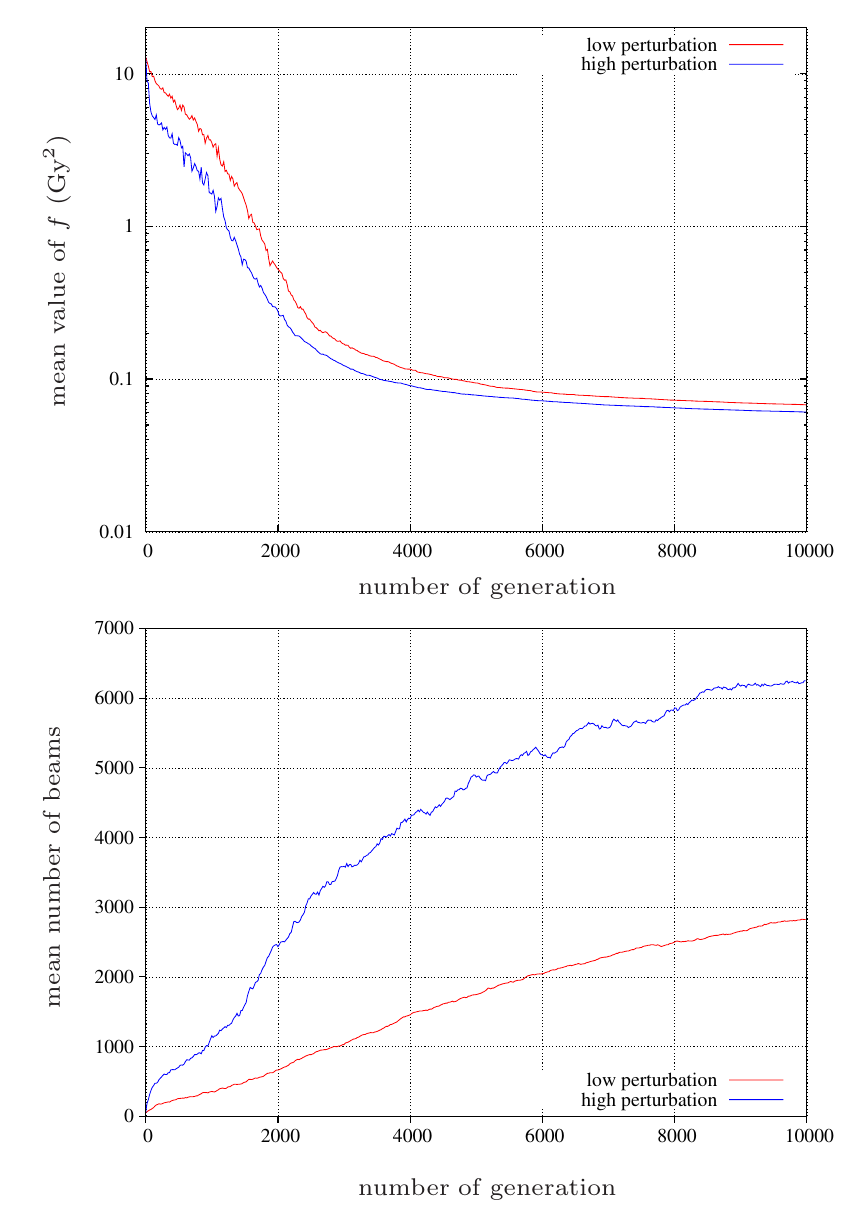}}
\caption[]{Influence of the crossover perturbation on the evolution. The convergence of the objective function (top) and the evolution of the mean number of coding beams (bottom) are represented for a $2$-beam (red lines) and $10$-beam (blue lines) FWHM Gaussian perturbation.}
\label{Fig_crossoverPerturbation}
\end{figure}

Two tests were carried out with Gaussian perturbation $\delta$ of FWHM respectively equal to $2$ and $10$ beams (see section~\ref{sssect:crossover}). The population size was set to $50$ individuals, the initial beam number to $50$ and the crossover rate to $0.5$. The convergence of the objective function and the corresponding evolution of the number of beams are represented in figure~\ref{Fig_crossoverPerturbation}.

The results show that the cut-and-splice operator directly drives the evolution of the number of beams: with a high crossover perturbation (blue line) the number of beams increases much faster than with a low perturbation.

However, the objective function reaches almost identical mean values: the search is mainly driven by the mutation operator and not by the cut-and-splice crossover. A high crossover perturbation favors a research based on the quantity of genes to the detriment of mutation-based optimization. In addition, the low perturbation optimization considerably reduces the calculation time since a smaller number of beams has to be simulated.

\paragraph{Influence of initial beam number}

\begin{figure}[p]
\centerline{\includegraphics[width=.8\textwidth]{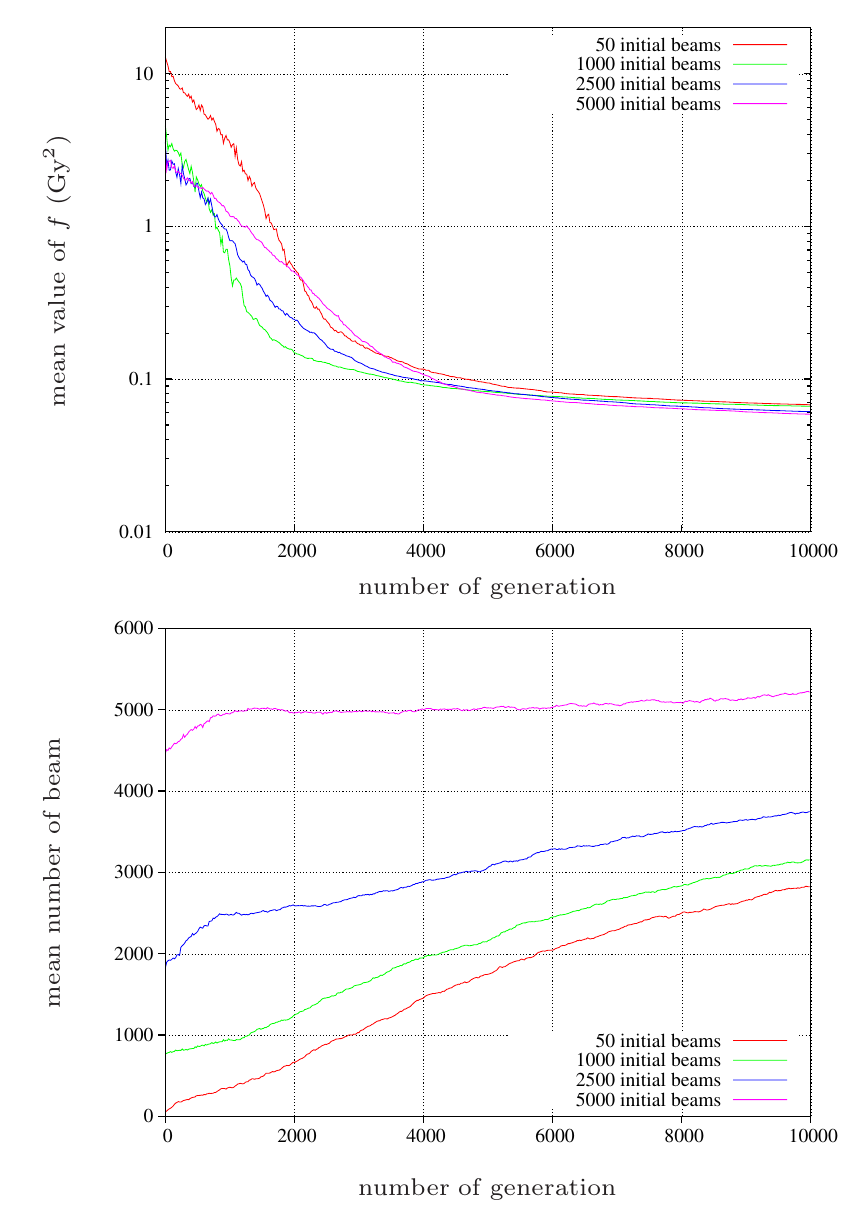}}
\caption[]{Influence of the initial beam number on the evolution process. The convergence of the objective function (top) and the evolution of the mean number of coding beams (bottom) are represented for $50$, $1000$, $2500$ and $5000$ initial beams.}
\label{Fig_initBeamNumber}
\end{figure}

The influence of the initial number of beams in the genetic process was studied by running tests with $50$, $1000$, $2500$ and $5000$ initial beams. The crossover perturbation was set to a FWHM of $2$ beams and all other genetic parameters were kept the same as in paragraph~\ref{par::crossoverPerturbation}. Figure~\ref{Fig_initBeamNumber} presents the convergence of the objective function and the corresponding evolution of the number of beams.

Like in the crossover perturbation study, the dynamics of the objective function are almost identical and converge to the same solution. The evolution of the number of beams for the different initial configurations seems to eventually tend to the same asymptotic value. However, the genetic optimization of the beam layout is more efficient when a smaller initial number of beams is considered: it is more difficult to optimize a large number of beams from scratch than to gradually integrate new beams at each generation.

\paragraph{Influence of fluence limit}

\begin{figure}[p]
\centerline{\includegraphics[width=.8\textwidth]{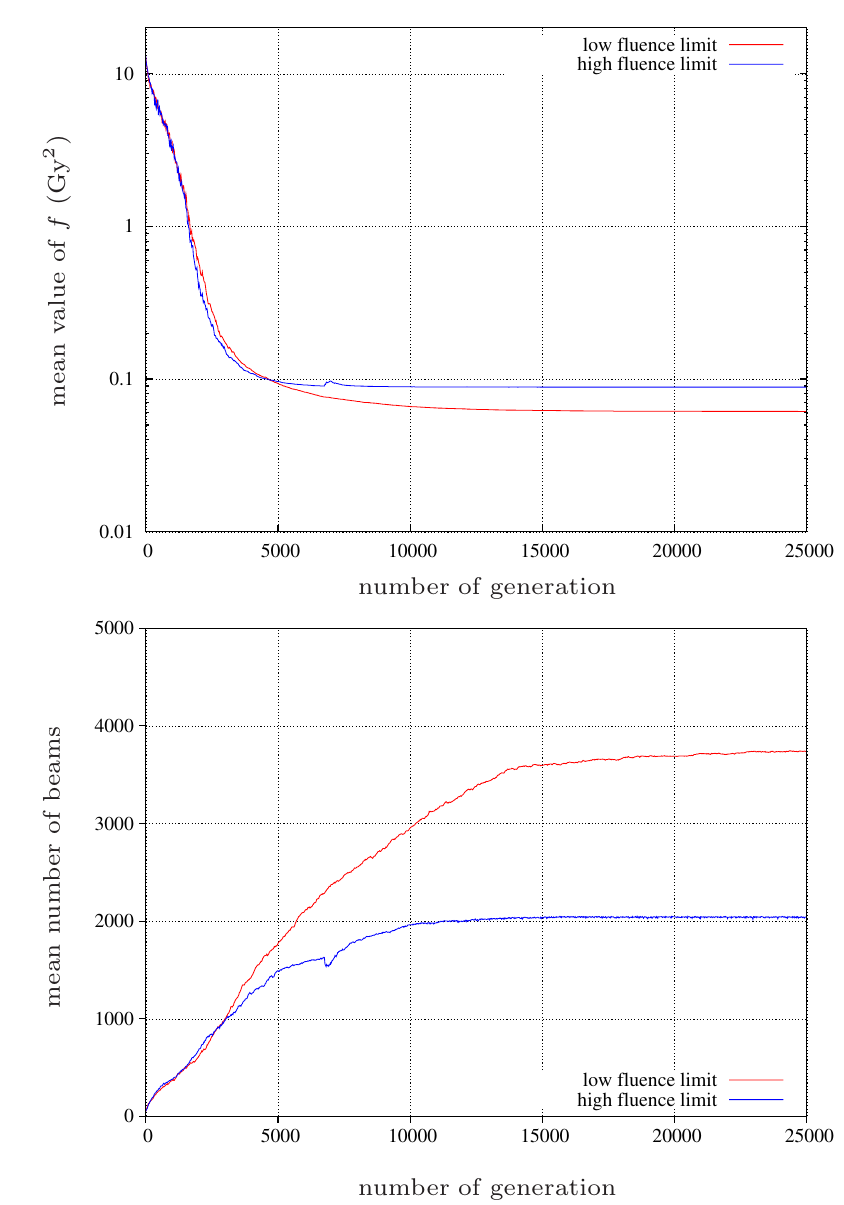}}
\caption[]{Influence of the fluence limit on the evolution. The convergence of the objective function (top) and the evolution of the mean number of coding beams (bottom) are represented for fluence limits of $2000$ and $10000$ particles per beam.}
\label{Fig_fluenceLimit}
\end{figure}

The role of the fluence limit is to prevent the formidable increase of the number of beams that would most likely happen with the cut-and-splice crossover. Two tests were carried out with fluence limits of $2000$ and $10000$ particles per beam and are represented in figure~\ref{Fig_fluenceLimit}.

Contrary to the crossover perturbation and initial number of beams, the fluence limit constrains the objective function as we can see in figure~\ref{Fig_fluenceLimit} (top). The optimization process displays two successive phases: (i) the number of beams increases first and the total number of incident protons (needed to obtain the prescribed dose in the PTV) is therefore gradually distributed over a larger set of beams, which reduces the average fluence per beam; (ii) then the fluence reaches the limit which stops the increase in the number of coding beams. The asymptotic value of the number of coding beams is directly related to the fluence limit, which is driven by the physics: a smooth depth-dose profile cannot be obtained with a poor statistics. We observed with the Geant4 MC code that a limit of $2000$ particles per beam leads to acceptable depth-dose profiles in the case of proton pencil beams. The quality of the beam layout is still given by the number of degrees of freedom (i.e.~the number of beams), the higher the fluence limit, the lower the quality of the beam layout at convergence.

\subsection{Clinically-realistic optimization}
\label{ssect:humanModelOptimization}

\begin{figure}[ht]
\centerline{\includegraphics[width=.7\textwidth]{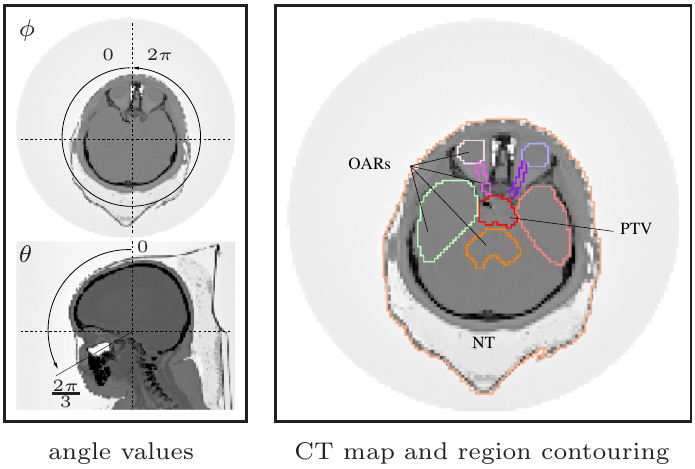}}
\caption[]{Human head model derived from a CT scan. The volume is composed of $128 \times 128 \times 107$ cubic voxels of edge $2.5$ mm. This model consists of a single PTV surrounded by the following OARs: the eyes, optic nerves, temporal lobes and brain stem.}
\label{Fig_humanModel}
\end{figure}

Our algorithm was tested on a clinically-realistic case based on a CT model of a human head (see figure~\ref{Fig_humanModel}). This model consists of a single PTV of $5.8$ cm$^3$ surrounded by typical OAR regions for a brain tumor case, i.e.~the eyes, the optic nerves, the temporal lobes and the brain stem. The rest of the head volume is considered as NT regions. The dose prescription to the PTV is set to $4$ Gy and the dose limits to OAR and NT regions respectively to $0$ and $0.01$ Gy. Note that for the eyes, the optic nerves and the temporal lobes, left and right parts are independent regions.

\begin{figure}[p]
\centerline{\includegraphics[width=.8\textwidth]{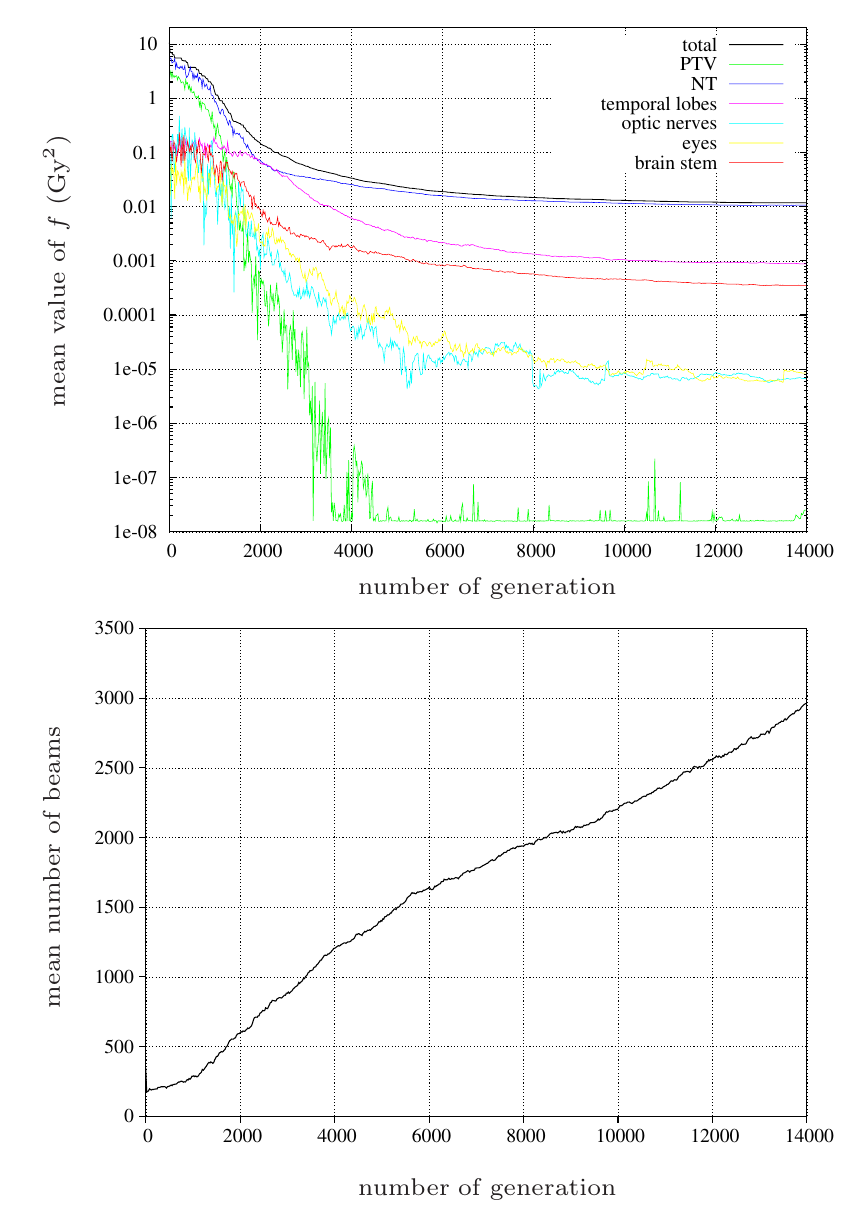}}
\caption[]{Convergence of the objective function and evolution of the mean number of coding beams for the human head model optimization. The total objective function as well as the contributions of each region are represented. For the sake of clarity, OAR regions of same type, e.g.~left and right eyes, were regrouped under a single label.}
\label{Fig_human_convergence}
\end{figure}

The optimization concerns the target positions and the beam incidence angles $\phi$ and $\theta$. Target positions $\textbf{p}$ are restricted to the PTV region, $\phi$ is defined between $0^\circ$ and $360^\circ$ and $\theta$ between $0^\circ$ and $120^\circ$ in order to avoid the irradiation of the inferior part of the body (figure~\ref{Fig_humanModel}, left). These parameters are randomly initialized. All the operators proposed in section~\ref{ssect:variableLengthGA} are used, with crossover rate and perturbation set to $0.5$ and $2$ beams FWHM, respectively. The population size is set to $50$ individuals and the initial number of beams to $200$. Figure~\ref{Fig_human_convergence} shows the convergence of the objective function components.

We can see that the relative contributions to the objective function evolve during the optimization. At the beginning, PTV and NT regions represent about $96 \%$ of the total objective function value. This is due to the low number of beams that cannot fully cover the PTV region and concentrate high dose in NT voxels. On the contrary, OAR regions benefit from this situation since only a few OAR voxels are irradiated. All along the optimization process, the contribution of the NT region remains the largest because every beam has to pass through NT. In spite of those unequal contributions, all the components of the objective function are optimized simultaneously. Note that the normalization of the region score by the number of voxels in the region (see section~\ref{sssect:objectiveFunctionAndSelection}) is the cause of higher noise in the convergence curve of small regions.

In the evolution of the number of beams (figure~\ref{Fig_human_convergence}, bottom), the slope changes after about $5600$ generations. Before this point, the PTV benefits from a fast increase in the number of beams, which leads to an excellent coverage of the tumor volume. Then the PTV objective reaches a minimum due to the stop condition of the gradient-based fluence optimization (see section~\ref{ssect:doseModel}). The optimization is then driven by the NT region for which the spreading of the dose and consequently a higher number of beams is useful.

\begin{figure}[t]
\centerline{\includegraphics[width=\textwidth]{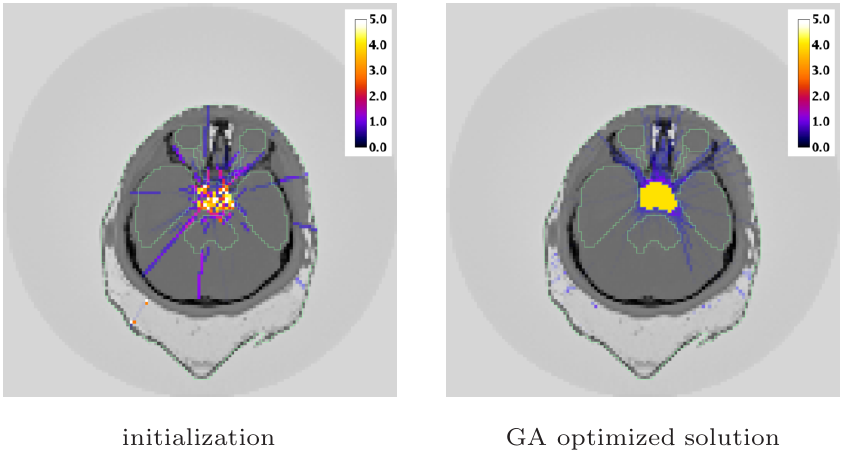}}
\caption[]{Initial and final dose maps ($50^{th}$ slice) of the brain tumor model optimization. For the sake of clarity, the contouring of regions has been overlaid to the dose maps and the dose outside the NT region has not been represented.}
\label{Fig_human_doseMaps}
\end{figure}

The initial and final dose maps are represented in figure~\ref{Fig_human_doseMaps} for the $50^{th}$ slice of the volume, which includes most OAR regions. In the initial dose map, the PTV coverage with $200$ beams is not satisfactory and a few beam trajectories pass through OAR regions. In the final dose map obtained with about $3000$ optimized beams, the dose in the PTV region is almost perfectly homogeneous (the largest deviation from the prescription is $7.5 \times 10^{-4}$). The algorithm has also found a suitable beam layout, avoiding OARs and limiting the dose to NT.

\section{Conclusion}

In this paper, we proposed a new inverse planning method intended to find the optimal beam layout for the spot scanning technique. A variable length GA with continuous parameters optimization combined with a gradient method is used to optimize simultaneously the beam number, the target positions and the incidence angles. This optimization tool was developped in a modular way in order to take into account multiple regions of different types, with different dose constraints. We took great care in choosing the genetic operators to make the best use of the GA characteristics:

\begin{itemize}
\item{} we adopted a non-uniform mutation scheme adapted to real coding which does not require the definition of a mutation rate. The genetic process is accelerated by a reinitialization of non-coding beams.
\item{} we introduced a modified cut-and-splice crossover operator to make the beam number evolve smoothly.
\item{} a rank-based selection operator is used to avoid premature convergence.
\end{itemize}

The dynamics of the proposed GA was first studied using a simple optimization test case. The influence of the initial beam number and mutation and crossover operator settings was discussed. It was found that simple rules are sufficient to tune the algorithm and obtain efficient convergence. Our method was then successfully used in a clinically-realistic test case.

\bibliographystyle{unsrt}
\bibliography{arxiv}

\end{document}